# 4DFlowNet: Super-Resolution 4D Flow MRI using Deep Learning and Computational Fluid Dynamics

Edward Ferdian[1], Avan Suinesiaputra[1,2], David Dubowitz[1], Debbie Zhao[3], Alan Wang[1,3], Brett Cowan[1,4], Alistair Young[1,5*]

[1] Department of Anatomy and Medical Imaging, University of Auckland, Auckland, New Zealand.

[2] School of Computing, University of Leeds, Leeds, United Kingdom

[3] Auckland Bioengineering Institute, University of Auckland, Auckland, New Zealand

[4] Institute of Environmental Science and Research, Auckland, New Zealand

[5] Department of Biomedical Engineering, King's College London, London, United Kingdom

*alistair.young@kcl.ac.uk

## Abstract

4D-flow magnetic resonance imaging (MRI) is an emerging imaging technique where spatiotemporal 3D blood velocity can be captured with full volumetric coverage in a single non-invasive examination. This enables qualitative and quantitative analysis of hemodynamic flow parameters of the heart and great vessels. An increase in the image resolution would provide more accuracy and allow better assessment of the blood flow, especially for patients with abnormal flows. However, this must be balanced with increasing imaging time. The recent success of deep learning in generating super resolution images shows promise for implementation in medical images. We utilized computational fluid dynamics simulations to generate fluid flow simulations and represent them as synthetic 4D flow MRI data. We built our training dataset to mimic actual 4D flow MRI data with its corresponding noise distribution. Our novel 4DFlowNet network was trained on this synthetic 4D flow data and was capable in producing noise-free super resolution 4D flow phase images with upsample factor of 2. We also tested the 4DFlowNet in actual 4D flow MR images of a phantom and normal volunteer data, and demonstrated comparable results with the actual flow rate measurements giving an absolute relative error of 0.6 to 5.8% and 1.1 to 3.8% in the phantom data and normal volunteer data, respectively.

## 1    Introduction

Cardiovascular magnetic resonance imaging (MRI) is a rapidly advancing non-invasive quantitative imaging method which enables precise evaluation of heart function. While being able to image the time-varying cardiac anatomy with high contrast, it can also acquire images of intravascular hemodynamics with blood velocity encoded in the phase of the MRI signal. Recent developments enable full 4D mapping (3 spatial dimensions plus time) of intravascular flow. 4D Flow provides a promising clinical utility to assess the hemodynamics of the blood inside the heart chambers and the great vessels for patients with cardiovascular disease [1]–[5].

Although 4D flow MRI provides complete coverage of blood flow inside the cardiovascular system, it still has limitations associated with signal-to-noise ratio (SNR), velocity encoding (VENC) and spatiotemporal resolution [6]. As the current resolution for 4D flow MRI is limited, some of the hemodynamic parameters, such as wall shear stress, cannot yet be calculated accurately.

To obtain improved resolution, several studies have explored the use of computational fluid dynamics (CFD) in combination with 4D flow MRI [7]–[10]. CFD simulations are computed by solving the continuity equation and Navier-Stokes equation within the region of interest. Compared to 4D flow MRI, CFD is able to achieve higher spatial and temporal resolutions. However, CFD solutions are dependent on accurate geometry as well as personalized inlet and outlet boundary conditions, in which 4D flow MRI can complement to a certain extent. In keeping with the ability of CFD to accurately model blood flow with (theoretically) unlimited spatiotemporal resolution, we took advantage of this to generate high resolution (HR) flow images, and model it as an image super resolution (SR) problem.

Recent advances in image super resolution using deep learning [11]–[13], have shown capabilities in enhancing image resolution, filling in missing details, and information recovery. However, image super resolution remains a challenging task and an ill-posed problem. Although advances in natural images and computer vision have lately been adopted for medical images [14], [15], none of these studies worked with velocity fields or 4D flow MRI representations (i.e. phase and magnitude images).

In this study, we propose a novel deep learning approach for super resolution network to increase the spatial resolution of 4D flow MRI, trained on purely synthetic 4D flow MR data. The synthetic 4D flow MR data were generated from CFD solutions and were made consistent with the image representations and physics of MRI. The deep learning network was trained to learn the mapping from noisy low resolution (LR) to noise-free HR phase images. To validate the method and test whether this mapping is also applicable to actual 4D flow Magnetic Resonance (MR) images, we evaluated our method with both synthetic and *in vitro* 4D flow MRI data in a flow phantom imaged at two resolutions, as well as an *in-vivo* scan of a normal volunteer.

## 2 Methods

### 2.1 Generating Training Data

The ideal training data set for this study would include large numbers of paired low resolution and high resolution 4D flow MRI images. However, collecting these data pairs is not economically feasible. Instead, we propose an approach to generate training data from CFD simulated flow data.

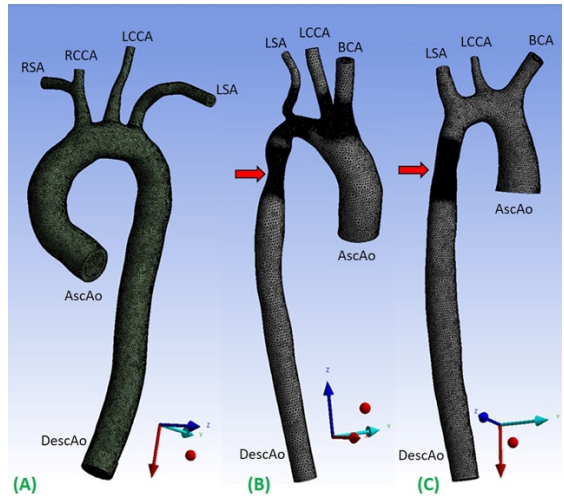

**Figure 1** Three aorta geometries for CFD simulations. Aorta01 (A) was extracted from PC-MRA of a normal volunteer data, Aorta02 (B) was modified from the MICCAI-STACOM 2012 CFD challenge, and Aorta03 (C) was taken from the MICCAI- STACOM 2013 CFD challenge. The red arrows mark the coarctation region. Inlet and outlet boundaries are shown in the Figure. Description: AscAo=ascending aorta, DescAo=descending aorta, RSA=right subclavian artery, RCCA= right common carotid artery, LCCA=left common carotid artery, and LSA=left subclavian artery, BCA=brachiocephalic artery.

We used three aortic geometries for the CFD simulations from a healthy volunteer (aorta01) and two data sets from MICCAI-STACOM CFD challenge in 2012 (aorta02) and 2013 (aorta03) [16], [17], which both have narrowing in the aortic vessel (coarctation). The aorta01 geometry was extracted from a 4D flow MRI study (spatial resolution 2.375 x 2.375 x 2.4 mm) by using temporal mean phase-contrast magnetic resonance angiogram (PC-MRA) surface extraction by using Paraview [18], and manual 3D geometry modeling by using

Blender 2.8 [19]. The rough geometry was automatically extracted using thresholding, connected component analysis, and surface extraction with triangulation in Paraview. Blender was then used to manually refine the initial rough geometry.

Computation meshes with tetrahedral elements were created for these three geometries by using ANSYS Meshing 19.2 from ANSYS Workbench [20]. Mesh convergence studies were performed on steady-state simulations for each of the geometries. ANSYS CFX 19.2 was used to run the CFD simulations.

We set the ascending aortic root as the inlet and the descending aorta as the outlet. Aorta01 had 4 aortic arch branches, right subclavian artery, right common carotid artery, left common carotid artery, and left subclavian artery. For aorta02 and aorta03, the first two branches were simplified by a common brachiocephalic artery. The three aortic geometries used for the CFD simulations are shown in **Figure 1**.

The following boundary conditions were used on each of the aortic geometries: velocity waveforms at the inlet, constant static pressure ($P_o = 0$ Pa) was assumed at the outlet, and pressure waveforms at the aortic arch branches. Velocity waveforms were obtained from the healthy volunteer data measurements using Siemens 4D Flow Demonstrator V2.4 (Siemens Healthineers, Erlangen, Germany) by placing 2D planes at each location (inlet, outlet, aortic arch branches). Due to limited spatial resolution, the measurements in the aortic branches were obtained from the root of the brachiocephalic artery (BCA) and left subclavian artery (LSA). Velocity, flowrates, and pressure were directly available from Siemens 4D Flow software. Pressure waveforms were calculated using the simplified Bernoulli equation within the software. The extracted pressure measurements were then recalculated relative to the constant pressure at the descending aorta.

In our CFD simulations, we prescribed a uniform velocity profile in the aortic inlet. Previous studies have shown that the inlet velocity profile has no significant impact on the flow region in the aorta, whereas the choice of outlet boundary condition affects larger regions of the flow in the aorta [21]. Even though our boundary conditions were not as realistic as some others, e.g. the Windkessel model, they were able to simulate reasonable aortic flow patterns suitable for training [22].

A time step of 0.01s was used for the transient simulation for one heart cycle, with a total time of 0.71s. The second order backward Euler transient scheme was employed. The Navier-Stokes equations were solved in ANSYS CFX 19.2. Average Reynolds numbers (Re) per heart cycle were 1170, 876, and 1170 for aorta01, aorta02, and aorta03 respectively; at the peak flow Re reached 4530, 3390, and 4530 respectively. However, most of the Reynolds number within the cycle were low (~1000), thus laminar flow was used to model the transient simulation with a maximum RMS residual of $10^{-4}$.

A no-slip boundary condition and rigid-wall assumption were applied on the vessel wall. Blood was modelled as a Newtonian fluid with a density of 1060 kg/m³ and viscosity of 4 .10⁻³ Pa s. The same simulation setup and boundary conditions were applied for all aortic geometries. Geometry properties, inlet and outlet boundary conditions are shown in **Table 1** and **Figure 2**.

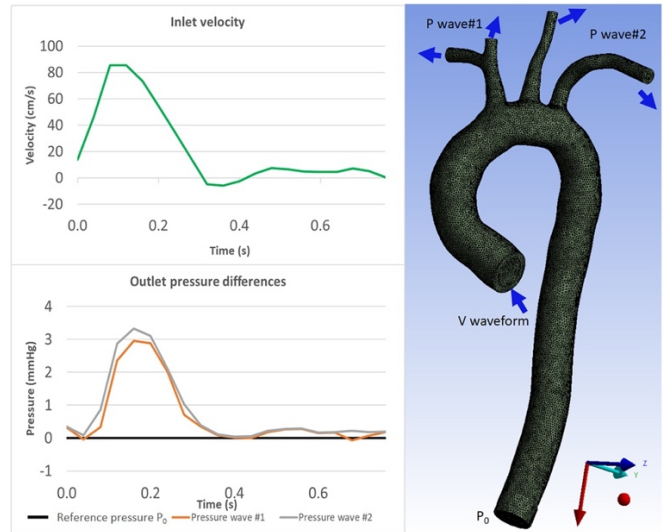

**Figure 2** Boundary conditions for CFD simulation: velocity waveform for inlet boundary condition (top) and pressure waveforms for the outlet boundary conditions (bottom) which were acquired from 4D Flow MRI measurements. The aorta geometry image (right) shows where the boundary conditions are prescribed. Inlet velocity waveform is prescribed as the inlet boundary condition; constant pressure $P_0$ is prescribed at the descending aorta as outlet boundary condition; and pressure waveforms are prescribed at the aortic branches as outlet boundary conditions (see **Table 1**).

Each CFD simulation resulted in 72 time frames, with the first frame omitted due to zero initialization values, resulting in a total of 71 frames. We extracted the velocity data from each time frame by using point clouds with uniform grid of 0.594 mm spacing projected into multiple planes with the same spacing. Velocity fields were represented as velocity vectors. For each time frame, three velocity images were obtained, representing three velocity components, $V_x$, $V_y$, and $V_z$, which correspond to the x, y, and z axes, respectively. These were then treated as the ground truth noise-free HR images.

### 2.1.1 Bridging the "CFD-4D Flow" gap – downsampling in the frequency domain

The generated velocity images from CFD simulations do not share the same characteristics with 4D flow MRI images. The velocity images are noise-free and not bounded to the value of the VENC parameter, which is an MR parameter to be specified prior the acquisition to adjust the maximum velocity corresponding to an 360° phase shift in the data. In 4D flow MRI, complex-data images are acquired according to the MRI point-spread function [23], and velocity information is encoded in the phase, whereas the magnitude is dependent on the transverse magnetization composition of the voxel (with signal components from fluid, soft tissue or air).

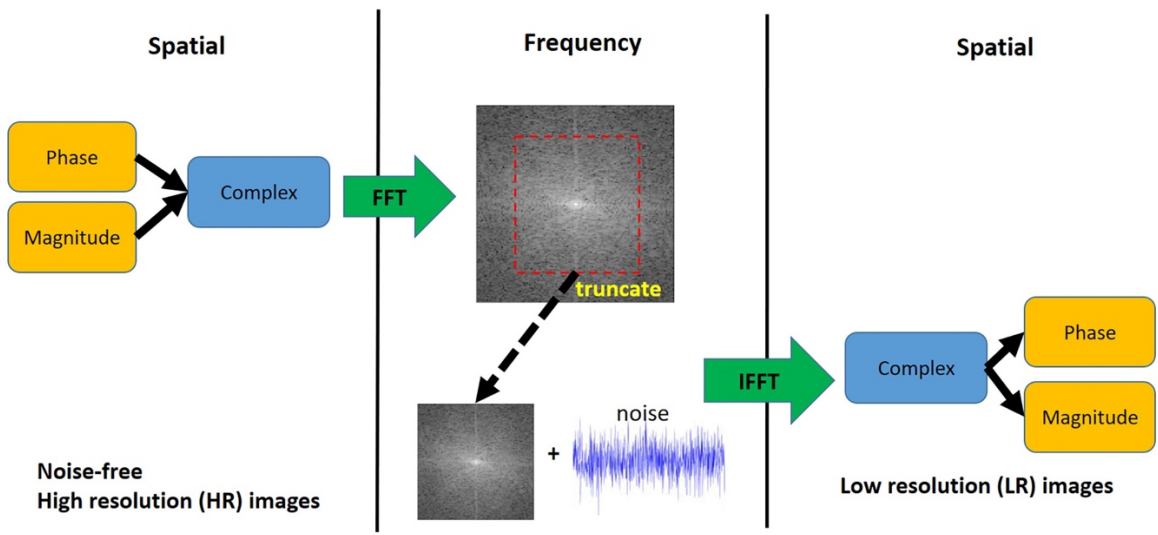

**Figure 3** Downsampling step for synthetic 4D flow phase images. White Gaussian noise was added in the frequency domain to mimic actual MRI acquisition. The visualization shows a 2D representation of the k-space while the actual downsampling step is performed in 3D.

In order to represent the velocity image as a 4D flow MRI equivalent, we performed the following steps for each velocity component image: 1) Choose a VENC higher than the maximum velocity in the image; 2) encode the velocity component into a phase image, within a range of $-\pi$ to $\pi$; 3) create a magnitude image with a non-zero constant value in the fluid region, representing the magnitude of the fluid signal, and a zero magnitude outside the flow, representing no-signal regions in which the phase is undetermined. Although actual 4D flow MR images also have static tissue (non-zero magnitude, low flow) adjacent to vessels, we found that low flow images were adequate for mimicking static tissue regions so we did not include extra static tissue regions in the simulations. Additionally, the network benefitted from the presence of no-signal regions in the training data, which improved the recognizability between the different characteristics of the noise in regions with and without signal.

To make sure the characteristics of 4D flow images were retained, we downsampled the images in the frequency (k-space) domain and added the corresponding noise to the LR images. The noise in the k-space is Gaussian for both real and imaginary signals [24]. Therefore, we added white Gaussian noise in the complex signal, resulting LR images with the appropriate noise distribution in the no-signal region, i.e. uniform noise distribution in the phase image and Rayleigh noise distribution in the magnitude image. The downsampling factor was 2.

The steps of the downsampling method (**Figure 3**) were as follows: 1) Compute the complex numbers from the phase and magnitude images. 2) Apply the fast Fourier transform to convert the complex numbers from spatial to frequency domain (k-space). 3) Truncate the outer-part (high frequency) information of the 3D k-space along the three axes so the dimension becomes half the original. 4) Add a zero-mean Gaussian noise with a certain standard deviation ($\sigma$) to the k-space to reach the target SNR, 5) apply the inverse Fourier transform to convert the k-space back to spatial domain. 6) Compute the magnitude and phase images from the complex numbers.

We approximated the power of signal $P_x = \frac{1}{N} \sum_{n=0}^{N} |x(n)|^2$, where $x(n) = x_r(n) + i \cdot x_i(n)$, where $x(n)$ is a complex number. Using the equation $SNR_{db} = 10 \log \frac{P_x}{P_n}$, we could calculate the power of noise $P_n$ which is the variance ($\sigma$). Since $x(n)$ is complex, a complex signal's noise with a sigma of $\sigma$ can be added directly to the complex k-space.

An example of resulting HR and LR phase images are shown in **Figure 4**. The synthetic HR phase image is noise-free while the LR phase image contains noise (uniform noise in non-fluid region and Gaussian noise in fluid region, dependent on VENC). By using this paired dataset, the SR network target is two-fold, improving the resolution by a factor of 2 and removing the noise in the phase image.

### 2.1.2 Data augmentation and patch generation

Data augmentation was performed in several parts of the data preparation process. In the downsampling process, there were three types of data augmentation. First, VENCs were chosen randomly from a subset of [30, 60, 100, 150, 200, 250, and 300 cm/s] for each velocity component. The chosen VENC was always higher than the maximum velocity component at each frame to avoid phase aliasing. Secondly, a constant intensity value was chosen randomly from a range of 60-240 intensity values for the fluid region in the magnitude image. Finally, we added different noise levels, by varying the target SNR between 14-17 decibels. These augmentations were chosen randomly and applied to each time frame.

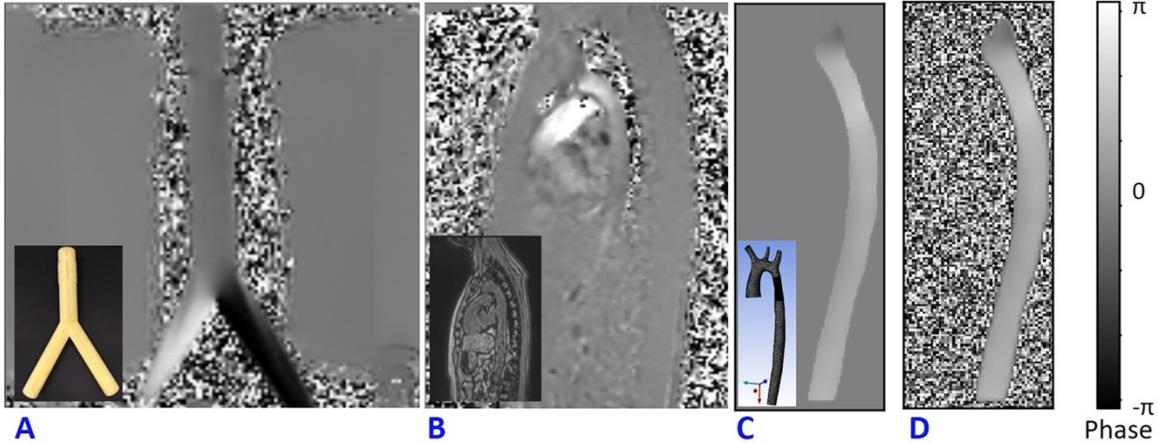

**Figure 4** An example of the different phase images. The figures shows the following: (A) phase image of actual 4D flow MRI of a bifurcation phantom; the real phantom is shown in the bottom left corner of the image. (B) Actual 4D flow phase image of a normal volunteer; the corresponding magnitude image is shown in the bottom left corner. (C) Noise-free synthetic phase image from CFD simulation result, which is treated as the ground truth high resolution phase image for the training dataset; the actual 3D geometry is shown in the bottom left corner. (D) Low resolution synthetic phase image with noise which is a result of the downsampling step. All values in the image are mapped between [-π, π] range representing negative and positive velocity values encoded as phase.

To compensate for the limited amount of shape and variations in the geometry, we selected a patch-based approach in training the super-resolution network. From each frame, we extracted 10 patches of 16x16x16 voxels (in the LR image). The locations of the patches were randomly selected, acting as random translations. For 9 out of 10 patches in each frame, we asserted a minimum fluid region of at least 20%, leaving the last patch unconstrained (potentially containing no fluid region). For every selected patch, we applied rotations in all 3 planes in 3 different angles, at 90◦, 180◦, and 270◦. As a result, 100 patches were obtained from each frame, adding up to a total of 7,100 patches per geometry.

The training set consisted of 14,200 patches from aorta01 and aorta02. Aorta03 was used for test and validation. The validation dataset consisted of 1 random patch per frame, with 9 different rotations, resulting in 10 patches per frame and a total of 710 patches. For the test dataset, patches from aorta03 were generated sequentially with a stride of ($n$-4), with $n$ as the LR patch size. Additionally, we utilized actual 4D flow MRI acquisitions of a bifurcation phantom and a healthy volunteer. The phantom dataset had two different isotropic resolutions (4mm and 2mm), while the healthy volunteer dataset has a single resolution (2.375 x 2.375 x 2.4mm).

## 2.2    Network architecture and Training

We developed a deep super resolution residual network (ResNet) called 4DFlowNet (**Figure 5**). The network was based on the generator part (SRResNet) of the SRGAN architecture [12]. We applied the upsampling layer using Tensorflow's bilinear resize function and utilized the residual blocks in both LR and HR space. The residual blocks in the LR space acted as a denoiser, while the residual blocks in the HR space refined the prediction after the upsampling layer. We used eight residual blocks before the upsampling layer and four residual blocks in HR space.

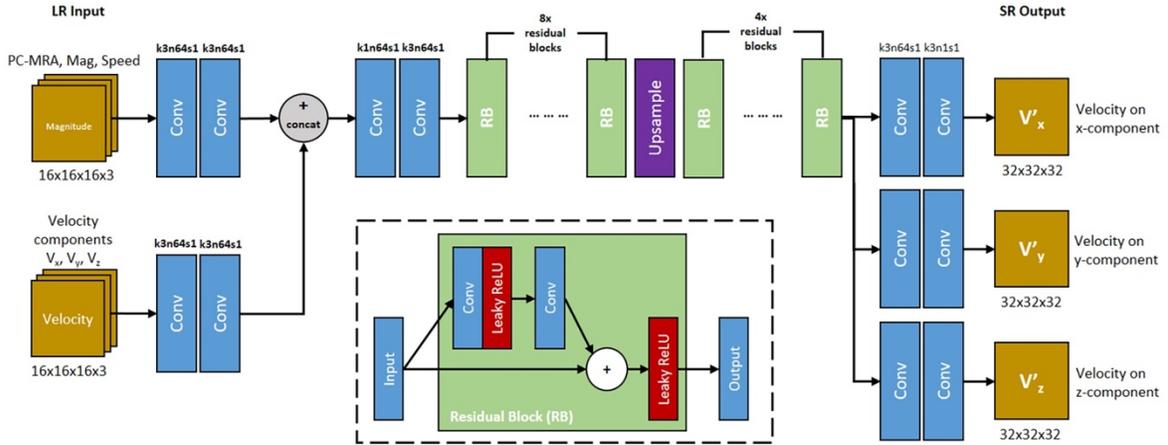

**Figure 5** 4DFlowNet architecture. The network utilizes 2 input path which represents anatomical information (top path) and velocity information (bottom path). RB represents the residual blocks. Conv represents 3D convolutions, with the number of kernel size, filters, and stride are shown above the operation. All convolution layers utilized symmetric padding and followed by a rectifier non-linearity (ReLU) except for output layers which used hyperbolic tangent (tanh) activation function. The network outputs 3 velocity components on its respective axis (V'$_x$, V'$_y$, V'$_z$). The inset shows the residual block. Convolution layer utilizes symmetric padding and 3x3x3 kernel. Plus (+) sign signifies the pixel-wise summation operation.

The input layers consisted of two separate paths, the anatomical path and the velocity path. Magnitude images were used as input in the anatomical path to help denoising the image and distinguishing fluid and non-fluid areas. The velocity path's input layer consisted of three channels, one for each velocity component ($V_x$, $V_y$, $V_z$). The anatomical images were composed of 3 channels: PC-MRA, speed, and mag channel. These images were calculated using the following:

$$Mag = \sqrt{M_x{}^2 + M_y{}^2 + M_z{}^2}$$

$$Speed = \sqrt{V_x{}^2 + V_y{}^2 + V_z{}^2}$$

$$PC - MRA = Mag * Speed$$

where $V$ represents the velocity, $M$ represents the magnitude, x, y, and z represent the velocity component on its respective axis. This means the main building block of the input layer consisted of 6 components: $V_x$, $V_y$, $V_z$, $M_x$, $M_y$, and $M_z$. While the network took the magnitude patches as input, the output consisted only the super-resolved velocity components ($V_x$, $V_y$, $V_z$).

The model took LR patches of 16 x 16 x 16 as input data and output 32 x 32 x 32 SR patches. Input and output velocity values were normalized to values within the range [-1, 1], with 1 being mapped to the maximum VENC of the phase image. In terms of images where different velocity components had different VENCs, the highest VENC was used to normalize the data. Magnitude values were normalized to values between [0, 1].

Before the output layers, the network branched into 3 separate prediction paths. Each velocity component was predicted in these separate layers instead of separate channels, in order to avoid shared-weighting between predicted velocity components [25].

One common characteristic of ResNet is its constant dimension throughout the different layers, which utilizes zero padding and affects prediction near the edge. This is normally not an issue for large image size, however for small patches, zero padding will corrupt the data and creates border artefacts. To avoid these border artefacts, symmetric padding was applied before performing every convolution. Each convolution layer was followed by a Rectified Linear Unit (ReLU) activation function. The output layer used a sigmoid tangent (tanh) activation function to make sure the output falls in the specified [-1, 1] range.

Our residual block consists of two convolution layers. The leaky ReLU activation function was utilized inside the residual block (**Figure 5Error! Reference source not found.**). Symmetric padding was also applied before each convolution layer in the residual block.

We implemented 4DFlowNet using Tensorflow 1.80 [26]. The Adam optimizer was used with the initial learning rate set to $10^{-4}$. A decay rate of $\sqrt{2}$ is applied to the learning rate after every 10,000 iterations. Due to memory constraints, we used a batch size of 20.

## 2.3 Loss function and accuracy metrics

The network was optimized using the mean squared error (MSE) and a weighted velocity gradient (VG) loss term. The pixel-wise MSE is applied to each velocity component, which reduces the magnitude velocity error. However, optimization by MSE tends to create blurry images, which affects velocity predictions near the vessel walls. To improve the quality of the image and prediction near the vessel walls, we introduced a velocity gradient loss term.

We formulated the loss function as the following:

$$l_{total} = l_{MSE} + 10^{-3} \, l_{VG}$$

The voxel-wise loss was calculated as:

$$l_{MSE} = \frac{1}{W \, H \, D} \sum_{x=1}^{W} \sum_{y=1}^{H} \sum_{z=1}^{D} (V'_x - V_x)^2 \; + (V'_y - V_y)^2 \; + (V'_z - V_z)^2$$

The velocity gradient loss was calculated as

$$l_{VG} = \frac{1}{W \, H \, D} \sum_{x=1}^{W} \sum_{y=1}^{H} \sum_{z=1}^{D} \left( \left( \frac{dV'_x}{dx} - \frac{dV_x}{dx} \right)^2 + \left( \frac{dV'_y}{dy} - \frac{dV_y}{dy} \right)^2 + \left( \frac{dV'_z}{dz} - \frac{dV_z}{dz} \right)^2 \right)$$

where first order differences were used to calculate the gradients: $\frac{dV_k}{dk} = \frac{V_{k+1} - V_{k-1}}{\Delta k}$ with $k \in \{x,y,z\}$. Here, $W$, $H$, and $D$ describe the dimensions of the output patch. This term helps the network in smoothing the gradient between the neighboring vectors and put more emphasis on the predictions near the vessel wall due to the high gradient values. Also, since incompressible fluid flow is theoretically divergence-free, and the CFD HR data is approximately divergence-free, this term helps the network learn to reproduce low divergence solutions.

### 2.3.1 Evaluation metric – relative speed error

Relative speed error was defined as the relative difference of velocity magnitude (speed) compared to the actual speed. To avoid division by zero, relative error is only evaluated in the fluid region. This was done by using a binary mask to exclude the non-fluid region. Additionally, we added a small number in the denominator ($\varepsilon=10^{-5}$) as a safety measure to avoid division by zero.

As a measure of accuracy, we used this metric to gauge the performance of the network. During training time, this metric was used to save the model checkpoint with the best relative speed error on the validation set.

$$rel\_err = \frac{1}{W\,H\,D}\sum_{x=1}^{W}\sum_{y=1}^{H}\sum_{z=1}^{D}\frac{\sqrt{\left(V'_x - V_x\right)^2 + \left(V'_y - V_y\right)^2 + \left(V'_z - V_z\right)^2}}{\sqrt{V_x^2 + V_y^2 + V_z^2} + \varepsilon}$$

## 3    Results

We performed our training on a Tesla K40 GPU with 12GB memory. Training took 10 seconds per iteration. The network was trained for over 100,000 iterations, which took approximately 10 days. Predictions were performed on image patches which were then stitched into a full volume. To perform the stitching method, LR patches were taken with a stride of ($n$-4) in each axis direction, with $n$ representing the patch size. During the image stitching process, 4-voxels were stripped from each border of the SR patches.

We tested 4DFlowNet on the following datasets: 71 time frames of aorta03 CFD (synthetic 4D flow MRI), 1 frame of bifurcation phantom data (comparing SR from 4mm to 2mm voxel size), and 1 frame of a 4D flow MRI volunteer dataset (treated as LR image with no HR data available).

To evaluate the performance of our method, we used relative speed error, average flow rate, and the divergence field. Due to availability of the HR images, we evaluated these metrics only for the synthetic 4D flow images from CFD simulations and 4D flow MRI phantom data. Additionally, we compared our result with linear interpolation, cubic spline interpolation, and sinc interpolation (i.e. by adding zero padding in k-space).

### 3.1    Tests on synthetic 4D flow MR images

**Figure 6** shows an example result of the network prediction on a patch of synthetic 4D flow MRI phase image compared to the ground truth and other interpolation methods. The figure shows a 2D slice from a 3D patch, with each velocity component shown separately in different rows. Additionally, we also show the divergence vector field of the patch. The prediction results were very close to the ground truth, while also substantially reducing the noise. On the other hand, the interpolation methods depended greatly on the noise level of the LR image and the relative velocity towards the VENC. It can be seen visually that the noise from the LR is interpolated to the SR image in both fluid and non-fluid regions.

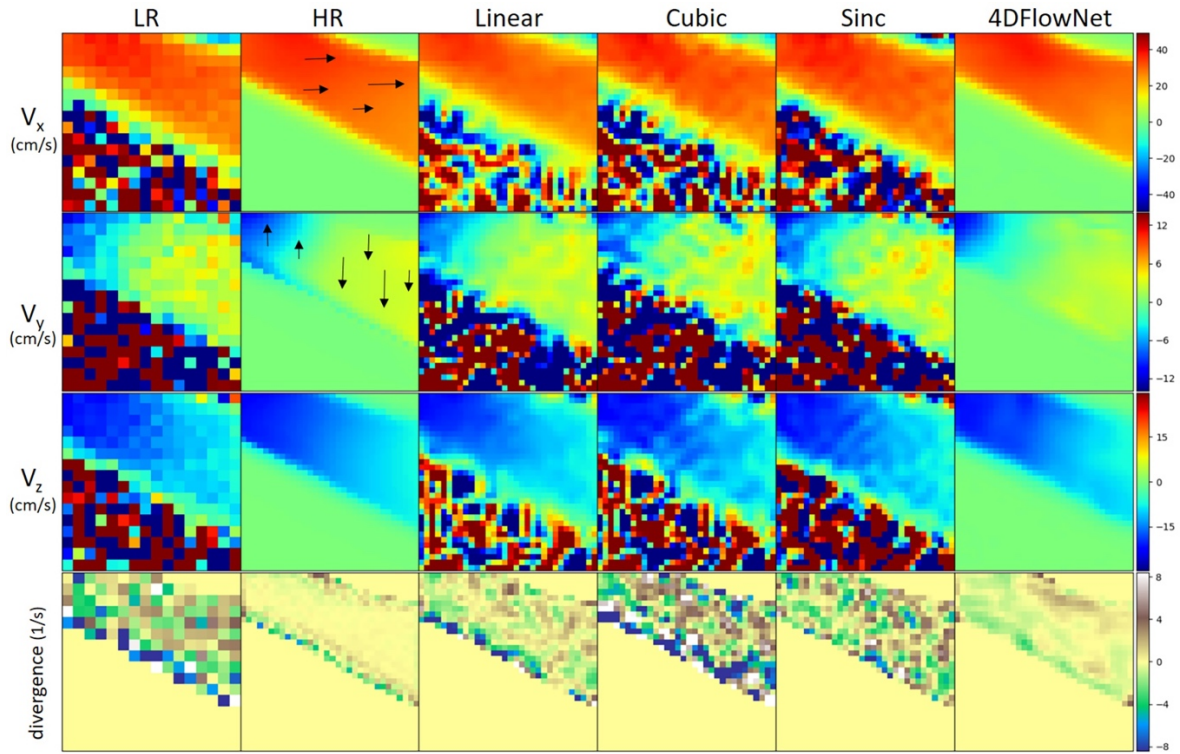

**Figure 6** Comparison of different upsampling methods applied to a patch from the synthetic 4D flow MRI phase image with different upsampling methods. Low resolution patch has a dimension of 16x16x16 and the upsampled patch has 32x32x32. For visualization, a 2D slice was taken from the patch. From left to right: low resolution (LR), high resolution (HR) / ground truth, linear interpolation, cubic spline interpolation, sinc interpolation, 4DFlowNet prediction. All velocity components have VENC of 100 cm/s. From 1st to 3rd row: velocity components on their respective x, y, z axis. Velocity scale was set to limits of dynamic range (in cm/s) for each of the velocity component. 4DFlowNet are robust in both high and low velocities, while other interpolation methods do not perform well in low velocity fields (i.e. $V_y$ and $V_z$). The fourth row shows a visualization of the divergence vector field of the respective patches. For better visualization, the divergence fields are only computed within the masked region (i.e. fluid domain).

In terms of divergence, we observed also that the network prediction produces smoother gradients and divergence fields closer to the ground truth. With the interpolation methods, the divergence fields were greatly affected by noise and non-uniformity of the vector fields.

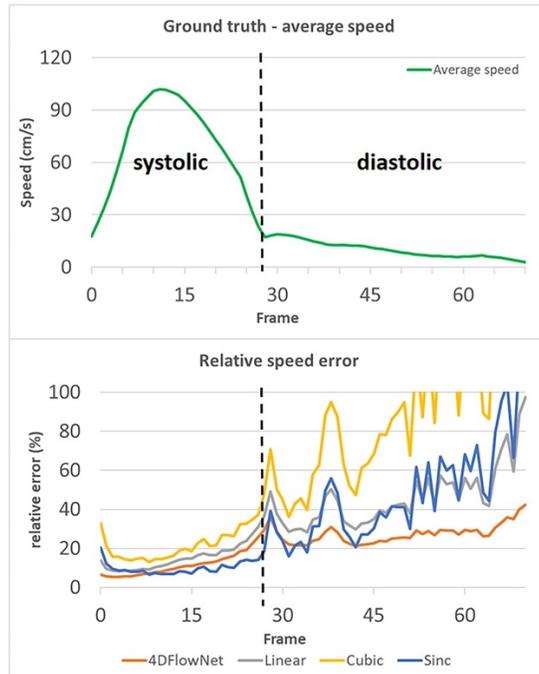

**Figure 7** Relative speed error of different methods compared to the actual ground truth speed, values over 100% are not shown. The relative speed error is calculated by taking an average of the voxel-wise relative speed error measurements within the masked fluid regions over the whole volume. Each time frame is considered as a different case. An increase in relative error occurs during the diastolic phase (frame 29-70), while the relative error remains low in the systolic phase (frame 0-28). The spikes in relative error occurred due to the actual ground truth speed having low values (relative to the VENC). 4DFlowNet is more robust towards prediction in low velocity fields, while other methods are less stable.

To measure the results quantitatively, we calculated the relative speed error (**Figure 7**), as measured in our evaluation metric. In this experiment, full volume predictions of 71 frames of the synthetic 4D flow images were utilized. Each of the frame was treated as a separate case. For a fair comparison with the other interpolation methods, the comparison was only performed within the fluid region. The relative speed error was calculated as an average of voxel-wise relative speed error between the methods compared to the ground truth speed. We observed comparable performance between the sinc interpolation and our network prediction in the systolic frames (frame 0-28), while our network achieved better performance for low velocity predictions during diastolic frames (frame 29-70). The relative error values were high for low velocity predictions due to the low relative velocity values (velocity values compared to VENC).

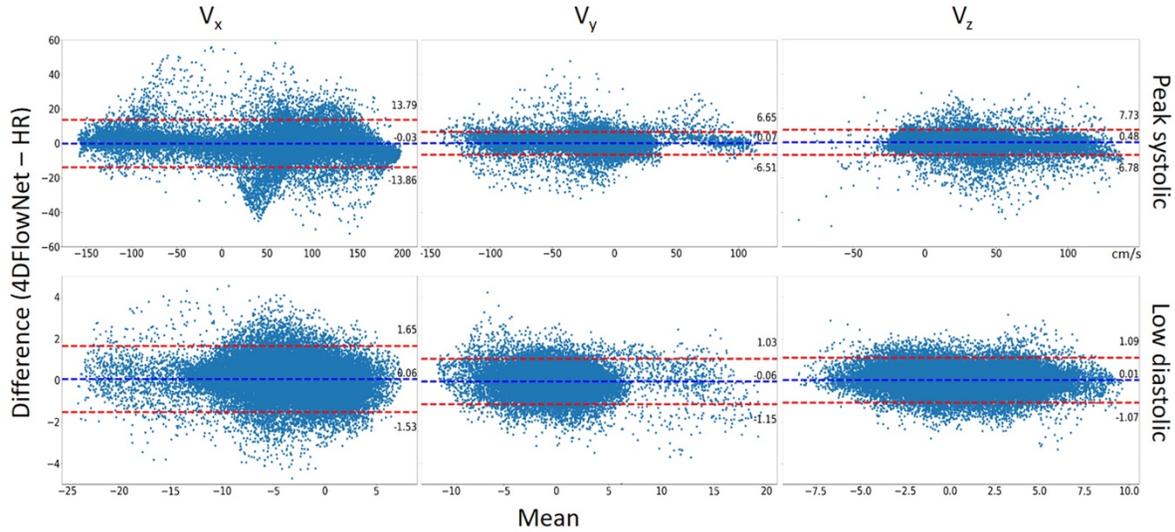

**Figure 8** Bland-Altman analysis. The comparison between 4DFlowNet predictions and ground truth was conducted on the $V_x$, $V_y$, $V_z$ velocity components for 50,000 random points within the fluid domain on peak systolic frame (top) and low diastolic frame (bottom). The dashed red lines indicate the 95% limits of agreement (mean ±1.96 * standard deviation).

Additionally, a Bland-Altman analysis was also performed to compare 4DFlowNet results to the ground truth HR at the peak systolic frame and one of the low diastolic frame for each velocity component ($V_x$, $V_y$, and $V_z$). For this analysis, 50,000 voxels were sampled randomly from within the masked fluid domain. Results in **Figure 8** show that the distribution of the error seem to be uniformly distributed around the mean.

Statistical tests (**Table 2**) showed that there were significant differences (p<0.05) for all the velocity components except for $V_x$ at peak systolic flow; however, this was likely due to the large number of data points since the bias in all cases was very low (<1% of peak velocity) and unlikely to be clinically significant. **Table 2** also shows the percentage relative speed error. It is likely that larger relative errors are due to lower velocity regimes, where a small change in prediction may cause relatively higher error. In the systolic frames, the main velocity component $V_x$ (see **Figure 1C**, aortic flow mainly exist in the x-axis) drives the flow. These results show that the 4DFlowNet performs well in both high velocity and low velocity regimes.

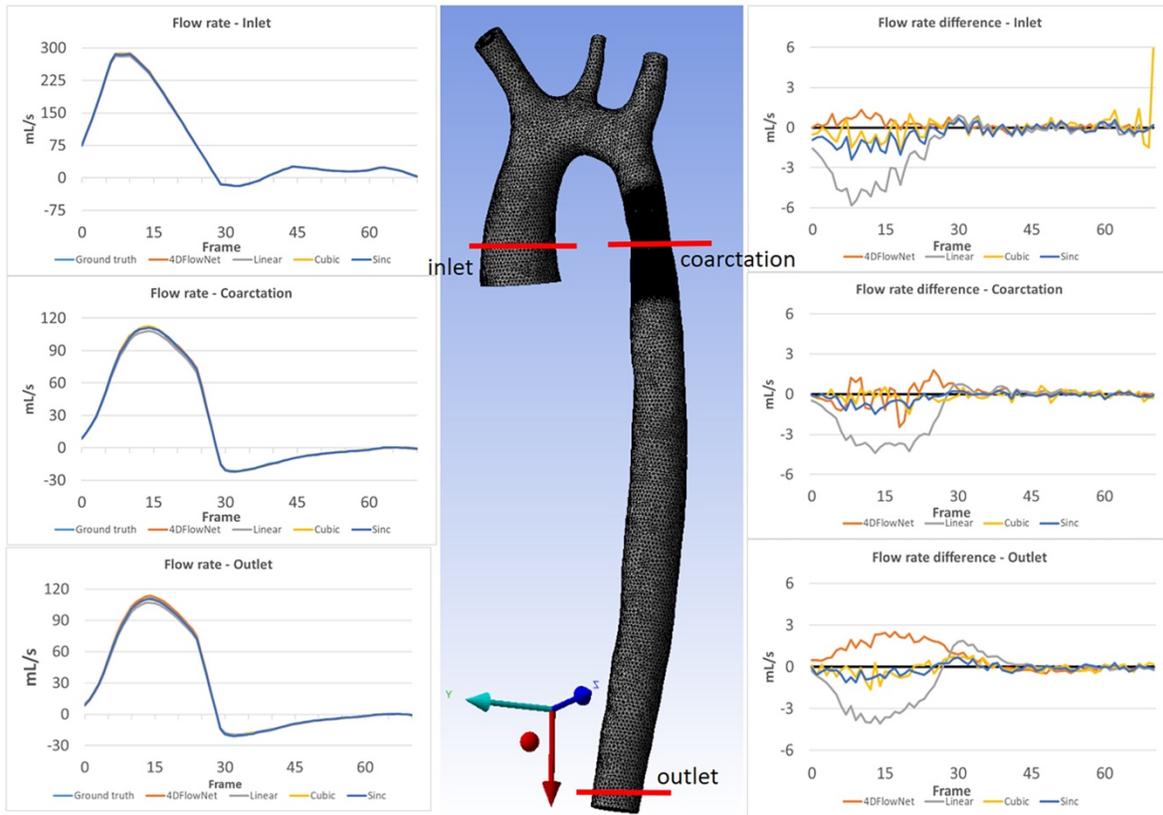

**Figure 9** Flow rate comparison at different slices: inlet, coarctation, and outlet for each time frame. The left column shows the flow rate comparison on 3 different slices for the ground truth and the different upsampling methods. The middle image shows the location of the analysis planes on aorta03. The right column shows the flow rate differences between the different methods compared to the ground truth flow rate.

Furthermore, we calculated the flow rate at three different cross-sectional planes on all the 71 time frames, as shown in **Figure 9**. Essentially flow rate is calculated from the integral of velocity vector going through the cross-sectional area, which will average out the noise. We observed comparable results and small differences between the 4DFlowNet predictions and the other interpolation methods. Consistent with the previous results, we observed higher error in the systolic frames, compared to the diastolic frames. These results indicate that the network is not introducing error in flow rates, an important clinical quantity.

### 3.2    Tests on actual 4D flow MRI data– bifurcation phantom

We also tested the network capability in predicting SR images from actual 4D flow MRI data. For this experiment, we utilized two different 4D Flow MRI resolutions with isotropic voxel size of 4mm and 2mm, respectively, in a flow phantom. We tested our network in predicting 2mm resolution (from 4mm resolution), and compared the prediction results with the actual acquisition at 2mm resolution.

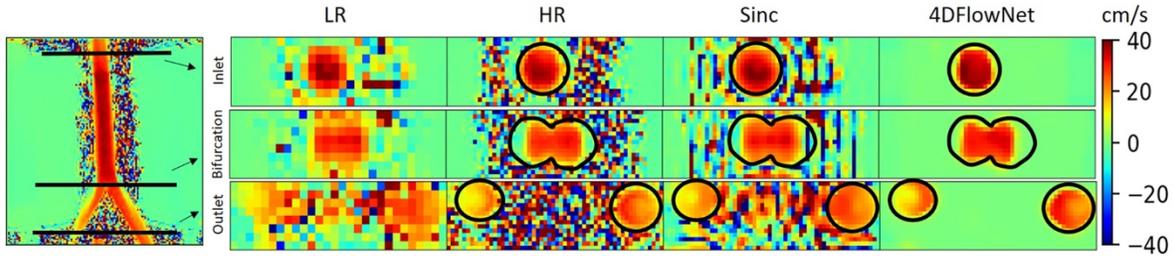

**Figure 10** Prediction result on actual 4D Flow MRI data of a bifurcation phantom. The most left image shows a 2D slice of the whole high resolution (HR) phantom data. Voxel sizes are 4mm for low resolution (LR) and 2mm for HR images. Three cross-sections are placed in the image to measure the velocity and flow rate comparison between the HR ground truth, sinc interpolation and 4DFlowNet predictions. The image shows the through-plane velocity. Flow rates (see **Table 3**) are calculated within the masked region (marked by black circles).

In this case, we only compared our results to the sinc interpolation method. Qualitatively the prediction result is shown in **Figure 10**. Similar to the result in synthetic 4D flow images, 4DFlowNet significantly reduces the noise outside the bifurcation phantom.

For quantitative analysis, three cross-section planes were taken at the inlet, bifurcation and the outlet of the phantom. The comparison between the flow rates for different analysis planes are shown in **Table 3**. Due to the noisy background, we prepared a binary mask to calculate the error only within the masked fluid region. In terms of difference in flow rate, 4DFlowNet offers slight improvement over the sinc interpolation method; the differences in relative error were -0.6% vs 7% at the inlet, 3.3% vs 4.3% at the bifurcation, and a comparable 5.8% at the outlet for 4DFlowNet and sinc interpolation, respectively. The flow rate measurements were compared to the acquired image to verify that the network is not adding any bias to the flow estimate.

### 3.3  Tests on actual 4D flow MRI data– normal volunteer data

To demonstrate the network performance in actual 4D flow MRI of a healthy volunteer (different from the one used for aorta01), we upsampled the acquisition resolution (2.375 x 2.375 x 2.4 mm) by a factor of 2, resulting in a resolution of (1.1875 x 1.1875 x 1.2 mm). This dataset was treated as LR image and HR ground truth image was not available. We performed this experiment to showcase the network's ability to enhance actual human 4D flow MRI data, while being trained exclusively on synthetic 4D flow MRI from aortic CFD data.

**Figure 11** shows the prediction results by the network, resulting in noise-free high resolution phase images. As a comparison, we provided the original LR counterpart as an inset in the subfigures. The reconstruction of the phase images were then visualized using Paraview [18]. **Figure 11** shows that most of the noise has been removed and the anatomy can be clearly seen. A streamline reconstruction was also performed to make sure there were no discontinuities introduced due to the image stitching from the patch-based approach. Border artefacts are not visible in the fully reconstructed volume and the stitching effect is seamless due to the convolution using symmetric padding.

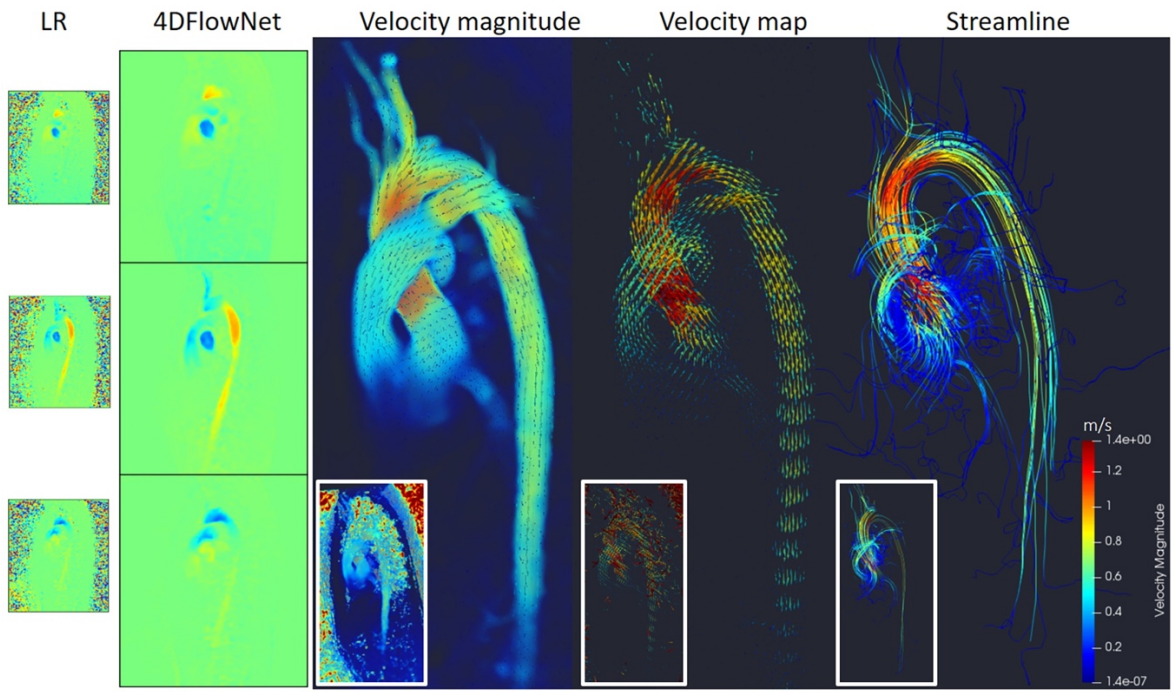

**Figure 11** Prediction results on actual 4D Flow MRI of a normal volunteer. The first column shows LR phase images, second column shows noise-free SR phase images predicted by the network. Visualization of the velocity magnitude (third image), velocity vector field (fourth image), and streamline reconstruction of the SR image (fifth image). The insets show the visualization of the LR images counterpart.

Additionally, we measured the flow rate on three cross-section planes during the peak flow. The result shows small flow rate differences: 10.7 mL/s (2.6%), -2 mL/s (-1.1%) and -4.8 (-3.8%) on the cross-sectional planes of ascending aorta, aortic arch, and descending aorta, respectively.

## 4    Discussion

We have developed 4DFlowNet, a deep super resolution residual network to increase the resolution of 4D flow MRI by using computational fluid dynamics as a proxy to generate training data. We demonstrated the potential of this application in actual 4D flow MRI of a phantom and healthy subject. Our flow predictions showed good agreement with the simulations, as well as phantom MRI acquisitions. Encouraging results were also obtained in a healthy volunteer.

Apart from providing high resolution phase images, our proposed network also removed noise in the phase images. Additionally, the network also improves the visibility of anatomical regions, which is influenced by the magnitude image patches that were incorporated in the input layers. As a result, the super resolution phase images contain clear boundaries between fluid and non-fluid regions. Better image quality is obtained with clear flow prediction near the vessel wall (**Figure 10**) and improvement towards divergence-free vector fields (**Figure 6**).

In our experiments, we have shown that the predicted velocity results have relatively small bias compared with ground truth CFD solutions. Additionally, 4DFlowNet produced smoother predictions compared to the other interpolation methods.

### 4.1    CFD as synthetic 4D flow MRI dataset

Our training dataset purely consisted of synthetic 4D flow images, which is not ideal compared to actual 4D flow MRI. Normally, with super-resolution problems, the LR data is obtained directly by downsampling original higher resolution image. Unfortunately, in 4D flow MRI, downsampling of the phase images would result in different noise distributions, which are then no longer representative of the actual phase data. Furthermore, the amount of high resolution 4D flow MRI data is also limited. By using synthetic 4D flow images, we indirectly address these limitations. Additionally, the use of data augmentation through different VENC and SNR also helps in adding variation to the training dataset, which otherwise is not economically possible with real MRI acquisitions.

While the CFD simulations were informed by real MRI measurements, we did not attempt to model accurate boundary conditions, non-linear viscosity, fluid-structure interactions or transitions to turbulence. Previous CFD studies [7]–[10], have been concerned with achieving accurate personalized CFD solutions given 4D flow data. In contrast, we have shown how CFD simulations can be used in a deep learning environment to learn how to reconstruct uncorrupted images from those corrupted by noise and low resolution. We therefore avoid the problem of accurately personalizing the CFD solution to a particular patient. Obtaining accurate CFD solutions is not a trivial task due to several factors, including image quality, unobtainable measurements in smaller vessels (i.e. aortic branches), and ability to obtain accurate pressure values. Some studies report significant differences between fluid-structure interactions and the rigid wall assumption in terms of wall shear stress [27], while other studies conclude that the rigid wall assumption is adequate and has little effect on the flow patterns [28], [29].

Also, only one cardiac cycle was simulated, which could give rise to transient initialization effects. However, negligible differences were observed when five cycles were simulated for one geometry (aorta01). While it may be beneficial to produce simulations (i.e. velocity fields) as close as possible to actual 4D flow measurements, it was found that the network was able to learn the filters necessary to enhance velocity fields, regardless of the flow profiles. Additionally, with our patch-based approach, the deep learning method is blinded to any kind of geometry or global flow information. We observed that the network was able to reconstruct velocity patterns in a phantom and in a volunteer data, despite being trained on relatively simple CFD simulations.

## 4.2 Network design

In designing the network, we took a patch-batch approach. Other than memory limitations, as well as mitigating the lack of training data due to limited geometrical variance, a patch-based approach also helped to obscure the contextual information. In our case, the synthetic phase data from CFD is different from the real acquired MRI phase data. The synthetic phase data has only fluid flow in the aortic geometry and consists of no other geometry, while actual 4D flow MRI phase image also contains flow information from other anatomical regions. The patch-based approach therefore helped to obscure any global information, while keeping only local information about the patch. While this approach may not be optimal, it generalises the network to a range of cases containing a fluid region. Furthermore, this could help the network in learning a variety of flow profiles, independent of the global geometry.

For the upsampling layer, we did not utilize the state-of-the-art approach, such as PixelShuffle [30] or sub-pixel convolution with nearest-neighbor initialization [31]. While these techniques were proven to be advantageous in recovering details and finer image textures in 2D super resolution networks [12], [32], we found they did not perform well for 3D velocity images. Additionally, the checkerboard artifacts were still prominent and the nearest-neighbor initialization did not solve the problem. As a result we utilized the conventional bilinear upsample layer, which required refinements in the HR space due to its blurry interpolated output.

During inference, 4DFlowNet accepts any arbitrary patch size (cube), up to the limit of the memory capacity. Inference time for an input size of 32 x 32 x 32 took roughly 1.2 seconds with a GPU, resulting in a 64 x 64 x 64 patch. Using 4DFlowNet, a full volume prediction and reconstruction (with image stitching) for a typical 4D flow MR image took 40-90 seconds, depending on the image size.

## 4.3 Limitations and Future Work

There are several limitations in our study in addition to the use of CFD simulations discussed above. One main limitation is the limited amount of geometry (three aortic geometries) and boundary conditions to generate our training dataset. While adding more aortic geometries can be beneficial for the training process, we aim to add different types of geometries in future work, such as ventricles and atria of the heart, so the network could learn more flow patterns. Additionally, different boundary conditions can be used to generate more training data with different flow profiles. Adding more training data with different geometries and different boundary conditions will enrich the capability of the network in distinguishing different flows. Also, the network should be validated on more 4D Flow MRI cases. In particular many patients with coarctation of the aorta have a bicuspid aortic valve and strongly helical aortic flow. Whether the current network can reproduce these flow features is a topic of future work. Finally, assumptions and optimizations which affect the estimation of wall shear stress should be investigated.

## 4.4 Conclusion

We have developed 4DFlowNet, a novel deep learning method of super resolution 4D flow MRI, which was trained solely on synthetic phase and magnitude images generated from CFD solutions. We have demonstrated the utility of this approach for actual 4D flow MRI data from phantom and normal anatomy. The results showed that the network provides a noise-free super resolution phase images with clear anatomical regions. This method has the potential to improve further from more training data, either synthetic or real 4D flow MRI data. The network was able to achieve the target upsampling factor of 2, and was also able to achieve a reduction in noise. The noise-free SR phase images can potentially be used to delineate regions of interest and automatically calculate flow parameters.

## 5     Conflict of Interest

Working expenses and partial stipend for EF was provided by Siemens Healthineers, Erlangen, Germany.

## 6     Author Contributions

All authors participated in conception and design of the study, interpretation of data, revision of the manuscript, and final approval of the submitted manuscript. EF wrote the first draft, designed the network architecture, and performed the data analysis.

## 7     Funding

This research has been funded by New Zealand Heart Foundation Scholarship, Grant No. 1786, the Health Research Council of New Zealand, programme grant 17/608, and a grant from Siemens Healthineers, Erlangen, Germany.

## 8     Acknowledgments

The authors wish to thank the funders.

**Table 1** Geometry properties, inlet and outlet boundary conditions for the three aorta geometries. Description: Re = Reynolds number, AscAo = ascending aorta, DescAo = descending aorta, RSA = right subclavian artery, RCCA = right common carotid artery, LCCA = left common carotid artery, and LSA = left subclavian artery, BCA=brachiocephalic artery.

| **Geometry** | **Aorta01** | **Aorta02** | **Aorta03** |
|---|---|---|---|
| Dimension (mm) | 232.8 x 83.4 x 105.9 | 32.3 x 52.3 x 167.4 | 203.3 x 48.5 x 61.2 |
| # elements | 0.65 x 10⁶ | 1.15 x 10⁶ | 0.99 x 10⁶ |
| **Inlet diameter** | 20 mm | 15 mm | 20 mm |
| **Re (average)** | 1170 | 876 | 1170 |
| **Re (peak)** | 4530 | 3390 | 4530 |
| **Boundary conditions** | | | |
| Inlet (AscAo) | Velocity waveform | Velocity waveform | Velocity waveform |
| Outlet (DescAo) | Constant pressure | Constant pressure | Constant pressure |
| BCA | Pressure waveform #1 | - | - |
| RSA | - | Pressure waveform #1 | Pressure waveform #1 |
| RCCA | - | Pressure waveform #1 | Pressure waveform #1 |
| LCCA | Pressure waveform #2 | Pressure waveform #2 | Pressure waveform #2 |
| LSA | Pressure waveform #2 | Pressure waveform #2 | Pressure waveform #2 |

**Table 2** Summary of prediction errors (mean ± standard deviation) of 4DFlowNet compared to the ground truth HR at peak systolic flow and a low flow diastolic frame of a synthetic 4D flow image. (*) indicates statistically significant differences between two measurements ($p < 0.05$). The results were sampled from 50,000 random points within the fluid domain. Peak velocity for each axis is also shown to give better context on the scale of error for each velocity component. Relative speed error was calculated as the voxel-wise difference in velocity magnitude (speed) compared to the actual speed.

|  |  | $V_x$ | $V_y$ | $V_z$ |
|---|---|---|---|---|
| **Peak systolic flow** | Prediction error (cm/s) | -0.03 ± 7.05 | 0.07 ± 3.36* | 0.48 ± 3.7* |
|  | Peak velocity (cm/s) | 200.46 | 144.95 | 143.89 |
|  | Relative speed error (%) | 7.05 ± 14.03 | | |
|  |  |  |  |  |
| **Low diastolic flow** | Prediction error (cm/s) | 0.06 ± 0.81* | -0.06 ± 0.56* | 0.01 ± 0.55* |
|  | Peak velocity (cm/s) | 23.84 | 20.53 | 9.88 |
|  | Relative speed error (%) | 23.16 ± 33.94 | | |

**Table 3** Comparison of flow rate of the bifurcation phantom 4D Flow MRI on two different resolutions. The flow rates are measured on three different planes (see **Figure 10**).

| Resolution LR -> HR | Slice location | Flow rate (mL/s) | | | Flow rate difference (mL/s) (% relative flow rate error) | |
|---|---|---|---|---|---|---|
| | | **Ground truth** | **4DFlowNet** | **Sinc** | **4DFlowNet** | **Sinc** |
| LR 4mm HR 2mm | Inlet | 111.6 | 110.9 | 119.4 | -0.7 (-0.6%) | 7.8 (7%) |
| | Bifurcation | 135.2 | 139.7 | 140.9 | 4.5 (3.3%) | 5.8 (4.3%) |
| | Outlet | 126.8 | 134.1 | 134.2 | 7.3 (5.8%) | 7.4 (5.8%) |